\documentclass[preprint,10pt]{elsarticle}

\usepackage{amsmath}
\usepackage{amssymb}
\usepackage{graphicx}
\usepackage{color} 
\usepackage{todonotes}

\begin{document}

\begin{frontmatter}
\title{Active nematodynamics on curved surfaces - the influence of geometric forces on motion patterns of topological defects}

\author[label1]{Michael Nestler}
\author[label1,label2,label3]{Axel Voigt}
\address[label1]{Institut f\"ur wissenschaftliches Rechnen, TU Dresden, 01062 Dresden, Germany}
\address[label2]{Center for Systems Biology Dresden (CSBD), Pfotenhauerstra{\ss}e 108, 01307 Dresden, Germany}
\address[label3]{Cluster of Excellence Physics of Life (PoL), 01062 Dresden, Germany}
\ead{axel.voigt@tu-dresden.de}

\begin{abstract}
We derive and numerically solve a surface active nematodynamics model. We validate the numerical approach on a sphere and analyse the influence of hydrodynamics on the oscillatory motion of topological defects. For ellipsoidal surfaces the influence of geometric forces on these motion patterns is addressed by taking into account the effects of intrinsic as well as extrinsic curvature contributions. The numerical experiments demonstrate the stronger coupling with geometric properties if extrinsic curvature contributions are present and provide a possibility to tune flow and defect motion by surface properties.
\end{abstract}


\end{frontmatter}

\newcommand{\wrt}{w.\,r.\,t.}
\newcommand{\eg}{e.\,g.}
\newcommand{\ie}{i.\,e.}
\newcommand{\cf}{cf.}
\newcommand{\oeda}{w.\,l.\,o.\,g.}
\newcommand{\etc}{etc.}

\newcommand{\formComma}{\,\text{,}}
\newcommand{\formPeriod}{\,\text{.}}

\newcommand{\domain}{\Omega}
\newcommand{\surf}{\mathcal{S}}
\newcommand{\surfDiscrete}{\surf_h}
\newcommand{\triangulation}{\mathcal{T}}
\newcommand{\feSpace}{\mathcal{V}(\surfDiscrete)}
\newcommand{\weak}[2]{\left( #1 \ , \ #2 \right)}
\newcommand{\dS}{\textup{d}\surf}
\newcommand{\R}{\mathbb{R}}

\newcommand{\testFctVec}{\boldsymbol{\psi}}
\newcommand{\testFctVecC}{\psi}

\newcommand{\testFctTen}{\boldsymbol{\Psi}}
\newcommand{\testFctTenC}{\Psi}

\newcommand{\tanPenP}{\omega_t^{\dirf}}
\newcommand{\tanPenQ}{\omega_t^{\qten}}

\newcommand{\pnorm}{\omega_{\text{n}}}

\newcommand{\ddt}{\frac{\textup{d}}{\textup{d}t}}
\newcommand{\dup}{\textup{d}}

\newcommand{\genvar}{\boldsymbol{q}}

\newcommand{\tb}{\boldsymbol{t}}

\newcommand{\para}{\boldsymbol{X}}
\newcommand{\dirfC}{p}
\newcommand{\dirf}{\boldsymbol{\dirfC}}
\newcommand{\DirfC}{P}
\newcommand{\Dirf}{\boldsymbol{\DirfC}}
\newcommand{\qtenC}{q}
\newcommand{\qten}{\boldsymbol{\qtenC}}
\newcommand{\QtenC}{Q}
\newcommand{\Qten}{\boldsymbol{\QtenC}}
\newcommand{\rtenC}{r}
\newcommand{\rten}{\boldsymbol{\rtenC}}
\newcommand{\stenC}{s}
\newcommand{\sten}{\boldsymbol{\stenC}}
\newcommand{\coords}{\boldsymbol{y}}
\newcommand{\vnor}{\mathcal{V}}
\newcommand{\normal}{\boldsymbol{\nu}}
\newcommand{\normalC}{\nu}
\newcommand{\conormal}{\boldsymbol{n}}
\newcommand{\orderp}{S}

\newcommand{\xvar}{\xi}
\newcommand{\dxvar}{\delta\xvar}

\newcommand{\genten}{\boldsymbol{r}}
\newcommand{\gentenC}{r}

\newcommand{\deformParam}{s}

\newcommand{\shop}{\boldsymbol{B}}
\newcommand{\shopC}{B}
\newcommand{\meanc}{\mathcal{H}}
\newcommand{\gaussc}{\mathcal{K}}
\newcommand{\g}{\boldsymbol{g}}
\newcommand{\LC}{\boldsymbol{E}}
\newcommand{\LCC}{E}

\newcommand{\proj}{\Pi}
\newcommand{\Id}{\operatorname{Id}}

\newcommand{\Tr}{\operatorname{Tr}}
\renewcommand{\div}{\operatorname{div}}
\newcommand{\Lie}{\mathcal{L}}

\newcommand{\dsurf}{\delta_{\surf}}

\newcommand{\bceq}{\overset{\text{BC!}}{=}}
\newcommand{\bcRightarrow}{\overset{\text{BC!}}{\Rightarrow}}

\newcommand{\tangent}{\operatorname{T}\!}
\newcommand{\tangentT}[1]{\tensor{\tangent}{#1}}
\newcommand{\qspace}{\mathcal{Q}\surf}

\newenvironment{fbalign}{\empheq[box=\fbox]{align}}{\endempheq}

\newcommand{\PotE}{\mathcal{U}}
\newcommand{\hatPotE}{{\hat{\PotE}}}
\newcommand{\hatPotEP}{\hatPotE^{\dirf}}
\newcommand{\hatPotEQ}{\hatPotE^{\Qten}}
\newcommand{\PotEP}{\PotE^{\dirf}}
\newcommand{\PotEQ}{\PotE^{\Qten}}
\newcommand{\PotEq}{\PotE^{\qten}}

\newcommand{\energy}{\mathcal{F}}
\newcommand{\helfrichEnergy}{\PotE_{\text{H}}}
\newcommand{\frankOseenEnergy}{\PotE_{\text{FO}}}
\newcommand{\landaudegennesEnergy}{\PotE_{\text{LdG}}}
\newcommand{\normalizationEnergy}{\PotE_{\text{n}}}
\newcommand{\bulkEnergy}{\PotE_{\text{B}}}
\newcommand{\elasticEnergy}{\PotE_{\text{El}}}
\newcommand{\areaEnergy}{\PotE_{\text{a}}}

\newcommand{\blueStart}{\color{blue}}
\newcommand{\blueEnd}{\color{black}}
\newcommand{\redStart}{\color{red}}
\newcommand{\redEnd}{\color{black}}

\newcommand{\helfrich}{Helfrich}
\newcommand{\frankOseen}{Frank-Oseen}
\newcommand{\landauDeGennes}{Landau-de Gennes}
\newcommand{\frankOseenHelfrich}{\frankOseen-\helfrich}
\newcommand{\landauDeGennesHelfrich}{\landauDeGennes-\helfrich}

\newcommand{\kinConst}{k}
\newcommand{\kinConstP}{\kinConst^{\dirf}}
\newcommand{\kinConstQ}{\kinConst^{\qten}}

\newcommand{\areaPen}{\omega_{\text{a}}}
\newcommand{\area}{A}
\newcommand{\areaZero}{\area_0}

\newcommand{\volumePen}{\omega_{\text{v}}}
\newcommand{\volume}{V}
\newcommand{\volumeZero}{\volume_0}

\newcommand{\topologicalCharge}{\Theta}
\newcommand{\trajectory}{\mathbf{d}}
\newcommand{\landau}{\mathcal{O}}

\newcommand{\xb}{\mathbf{x}}

\newcommand{\inputTikzPic}[1]{\ifthenelse{\boolean{plotTikzPics}}{\input{#1}}{\fbox{\centering\begin{minipage}[t][5cm][t]{0.9\textwidth}\centering\textbf{\color{red}enable plotTikzPics}\end{minipage}}}}

% differential operators
\newcommand{\Grad}{\nabla}
\newcommand{\Div}{\operatorname{div}}%
\newcommand{\Rot}{\operatorname{rot}}%
\newcommand{\BigRot}{\operatorname{Rot}}%
\newcommand{\DivSurf}{\Div_{\!\surf}}%
\newcommand{\GradSurf}{\Grad_{\!\surf}}
\newcommand{\RotSurf}{\Rot_{\surf}}
\newcommand{\BigRotSurf}{\BigRot_{\surf}}
\newcommand{\GuentherGradSurf}{\mathcal{D}}
\newcommand{\vecNabla}{\boldsymbol{\nabla}}
\newcommand{\levicivitaPlain}{\vecNabla}
\newcommand{\levicivita}[2]{\vecNabla_{\!\!#1}{#2}}
\newcommand{\levicivitaNondim}[2]{{\tilde\vecNabla}_{\!\!#1}{#2}}
\newcommand{\laplaceBeltrami}{\Delta_{\surf}}
\newcommand{\vecLaplace}{\boldsymbol{\Delta}}
\newcommand{\laplaceDeRham}{\vecLaplace^{\!\textup{dR}}}
\newcommand{\laplaceRotRot}{\vecLaplace^{\!\textup{RR}}}
\newcommand{\laplaceGradDiv}{\vecLaplace^{\!\textup{GD}}}
\newcommand{\laplaceDivGrad}{\vecLaplace^{\!\textup{DG}}}
\newcommand{\laplaceBochner}{\vecLaplace^{\!\textup{B}}}
\newcommand{\dotGradSurf}{\dot{\Grad}_{\!\surf}}

\newcommand{\upperdot}[1]{\overset{\scriptscriptstyle\blacktriangle}{#1}}
\newcommand{\lowerdot}[1]{\overset{\scriptscriptstyle\blacktriangledown}{#1}}

\newcommand{\functionalDerivative}[2]{\frac{\delta{#1}}{\delta{#2}}}
\newcommand{\materialDerivative}[1]{\dot{#1}}
\newcommand{\tangentialAcceleration}{{\boldsymbol{a}_m}}
\newcommand{\normalAcceleration}{{a_{m,\perp}}}

\newcommand{\metric}{{\boldsymbol{g}}}
\newcommand{\trace}{\operatorname{tr}_2}%

\newcommand{\surfaceTensorProduct}{{{:}_{\metric}}}
\newcommand{\stress}{\boldsymbol{\sigma}}
\newcommand{\tangentVelocity}{\boldsymbol{v}}
\newcommand{\tangentVelocityC}{v}
\newcommand{\Velocity}{\boldsymbol{V}}
\newcommand{\VelocityCmp}{V}
\newcommand{\helfrichConstant}{\alpha}
\newcommand{\landauDeGennesConstant}{L}
\newcommand{\abbrevH}{w_H}
\newcommand{\surfaceInnerProduct}[2]{\left\langle #1 , #2 \right\rangle_\metric}

\newcommand{\abbrevBulk}{w_{\textup{bulk}}}
\newcommand{\abbrevOne}{\boldsymbol{w}_1}
\newcommand{\abbrevZero}{w_0}

\newcommand{\relativeVelocity}{\boldsymbol{u}}
\newcommand{\abbrevBm}{{\boldsymbol{b}_{m}}}
\newcommand{\dynamicViscosity}{\nu}
\newcommand{\pressure}{p}
\newcommand{\surfaceId}{\operatorname{Id}_\surf}

\newcommand{\ts}{{t^*}}
\newcommand{\ls}{{l^*}}
\newcommand{\ms}{{m^*}}

\newcommand{\densityNondim}{\bar\density}
\newcommand{\kinConstQNondim}{\tilde\kinConstQ}

\newcommand{\f}{\boldsymbol{f}}
\newcommand{\abbrevG}{g}
\newcommand{\abbrevGE}{g_{\textup{E}}}
\newcommand{\abbrevGH}{g_{\textup{H}}}
\newcommand{\abbrevW}{\boldsymbol{w}}
\newcommand{\helfrichConstantGeneral}{\bar\helfrichConstant}
\newcommand{\Be}{\textup{Be}}
\newcommand{\Pa}{\textup{Pa}}

\newcommand{\intermediateVelocity}{\tangentVelocity^\star}
\newcommand{\deformationTensor}{\boldsymbol{d}}
\newcommand{\vorticityTensor}{\boldsymbol{\omega}}
\newcommand{\molecularField}{\boldsymbol{h}}
\newcommand{\activity}{\alpha}
\newcommand{\stressScale}{\Sigma}
\newcommand{\us}{{u^*}}

\newcommand{\density}{\rho}
\newcommand{\friction}{\gamma}
\newcommand{\viscosity}{\eta}

\newcommand{\unit}[1]{\left[ \mathrm{#1} \right]}
\newcommand{\ReynoldsNumber}{\mathrm{Re}}
\newcommand{\EricksenNumber}{\mathrm{Er}}
\newcommand{\ActivityNumber}{\mathrm{Ac}}
\renewcommand{\Re}{\ReynoldsNumber}

%Volume Versions
\newcommand{\DeformationTensor}{\boldsymbol{D}}
\newcommand{\VorticityTensor}{\boldsymbol\Omega}
\newcommand{\DeformationTensorCmp}{D}
\newcommand{\VorticityTensorCmp}{\Omega}

\newcommand{\MolecularField}{\boldsymbol{H}}
\newcommand{\traceR}{\operatorname{Tr}_3}%
\newcommand{\Pressure}{P}

%%%% Start %%%%%%
\section{Introduction}
\label{sec1}

The driving force behind the huge interest in collective behaviour of active matter is the goal to understand the physics of natural materials. One well-studied class of active matter, which includes, for example, epithelia cells, elongated bacteria and filamentous particles inside living cells, can be described by the interaction of rod-shaped particles. This relates these systems to nematic liquid crystals with long-range orientational order between these particles. 
Adapting these theories and extending them by active components leads to the concept of ‘active nematics’, see \cite{Doostmohammadi_NC_2018} for a review. The active contribution drives the system out of equilibrium and leads to spontaneous generation/annihilation of topological defects, destruction of long-ranged nematic order and the formation of active turbulence. 

If such systems are confined on curved surfaces, topological constraints strongly influence the emerging spatiotemporal patterns. Using these topological constraints to guide collective cell behavior might be a key in morphogenesis \cite{Doostmohammadi_PRL_2016} and active nematic films on surfaces have been proposed as a promising road to engineer synthetic materials that mimic living organisms \cite{keber2014topology}. As in passive systems the mathematical Poincaré-Hopf theorem forces topological defects to be present in the nematic film. On a sphere this leads to an equilibrium defect configuration with four +1/2 disclinations arranged as a tetrahedron \cite{Lubenskyetal_JPII_1992}. The disclinations repel each other and this arrangement maximises their distance. In active systems unbalanced stresses drive this configuration out of equilibrium. But in contrast to planar active nematics with continuous creation and annihilation of defects the creation of additional defect pairs can be suppressed on curved surfaces, which is demonstrated in \cite{keber2014topology} for an active nematic film of microtubules and molecular motors, encapsulated within a spherical lipid vesicle. This provides an unique way to study the dynamics of the four defects in a controlled manner and led to the discovery of a tunable periodic state that oscillates between the tetrahedral and a planar defect configuration. %However, in \cite{keber2014topology} these oscillations are slightly perturbed and the perturbations are speculated to result from hydrodynamic interactions.

Various modeling approaches have been proposed to describe the periodic defect motion. They range from a coarse-grained model in which the +1/2 disclinations are effectively described by self-propelled particles with a velocity proportional to the activity \cite{Giomi_PRL_2013}. On a sphere this approach leads to oscillations between the planar and tetrahedral configuration \cite{keber2014topology}. However, a quantitative comparison with the experimental results leads to differences, which become more evident for more general surfaces. For non-constant Gaussian curvature constraints local geometric properties influence the position of the defects and thus can be used to control defect dynamics. These effects are addressed with particle simulations \cite{alaimo2017curvature,Apaza_SM_2018,Ellis_NP_2018}. We here consider a continuous description and also account for hydrodynamic effects. The considered model belongs to the class of 'active nematodynamics', it is a simplified Beris-Edwards model with active driving, see \cite{julicher2018hydrodynamic} for a review. We propose a thin-film limit of this modeling approach and numerically solve the corresponding surface model. Related models have been considered in \cite{Torres-Sanchez_JFM_2019,Pearce_PRL_2019}. However, these models are based on a simplified surface Landau-de Gennes energy neglecting various curvature contributions \cite{Kralj_SM_2011}. For more detailed surface Landau-de Gennes models which also take extrinsic curvature contributions into account, see \cite{Golovatyetal_JNS_2017,Nitschke_PRSA_2018,Nestler_SM_2020}. Another critical issue is the considered numerical approach for the Navier-Stokes-like equations. In \cite{Torres-Sanchez_JFM_2019,Pearce_PRL_2019} it is based on a vorticity-stream function formulation and thus is restricted to surfaces which are topologically equivalent to a sphere \cite{Nitschke_PAMM_2021}. More general numerical approaches have been proposed in \cite{Nitschkeetal_book_2017,Reuther_PF_2018,Reuther_JFM_2020}. We here combine such a general formulation with a numerical approach for a surface Landau-de Gennes model with intrinsic and extrinsic curvature contributions \cite{Nitschke_PRSA_2018,Nitschke_PRSA_2020} and demonstrate the relation between flow, topological defects and geometric properties of the surface for surface active nematodynamics on surfaces with varying Gaussian curvature.   

The paper is structured as follows: Starting from the corresponding model in three dimensions we derive the surface model as a thin film limit in Section \ref{sec2}. The numerical approach to solve the surface active nematodynamics model is described in Section \ref{sec3}. We also provide convergence results of the surface finite element approach for the coupled system of vector- and tensor-valued surface partial differential equations. Results and numerical experiments are discussed in Section \ref{sec4} and conclusions are drawn in Section \ref{sec5}. Technical details are provided in the Appendices.

\section{Model}
\label{sec2}
%\paragraph{Active nematodynamics} 
Our starting point are the hydrodynamic modeling approaches for active nematodynamics in flat space, see \cite{giomi2014defect, thampi2014instabilities, julicher2018hydrodynamic}. These models are closely related to the classical Beris-Edwards model \cite{beris1994thermodynamics} but vary in the details of assumed transport type for the tensorial field \cite{nitschke2020observer}, as well as variational spaces or the definition of the molecular field. However, all models share the fundamental approach of coupling momentum and mass balances with a dynamic equation for nematic ordering driven by a gradient flow \wrt $\;$ to a free energy. We will base our modeling on \cite{julicher2018hydrodynamic,giomi2015geometry}, which provide congruent approaches and read in a three-dimensional space 
\begin{align}
    \density \frac{D\, \Velocity}{D\, t} = \nabla \cdot \stress ,\qquad \nabla \cdot \Velocity=0 ,\qquad \frac{D\, \Qten}{D \,t} + \VorticityTensor\Qten - \Qten \VorticityTensor = -\lambda \DeformationTensor + \Gamma^{-1} \MolecularField \label{eq:stateEqu}
\end{align}
with $\Velocity$ velocity, subject to incompressibility, $\stress$ stress and $\Qten$ the nematic order parameter, a symmetric trace-free second order tensor.
Here $D\, / D\, t = \partial_t + \Velocity \cdot \nabla $ denotes the material derivative and $\DeformationTensor = 1/2 (\nabla \Velocity + \nabla \Velocity^T )$ and $\VorticityTensor =1/2 (\nabla \Velocity - \nabla \Velocity^T ) $ are symmetric and antisymmetic contributions of the rate of deformation tensor. $\lambda$ denotes the flow alignment parameter for a reactive coupling between kinetic and Landau-de Gennes energy and $\Gamma^{-1}$ is a rotational viscosity. The molecular field $\MolecularField = - \delta \energy / \delta \Qten$ is defined along the variation, in space of trace-free symmetric tensors, of a free energy 
\begin{align*}
    \energy (\Velocity, \Qten) = \int_{\volume} \frac{\density \Velocity^2}{2} \mathrm{d}\volume + \int_{\volume} \frac{L}{2} \| \nabla \Qten \|^2 + a \traceR \Qten^2 + \frac{2}{3} b\traceR \Qten^3 + c \traceR \Qten^4 \mathrm{d}\volume
\end{align*}
consisting of a kinetic contribution and a one-constant approximation of the Landau-de Gennes energy \cite{virga1995variational} with $L$ and $a,\,b,\,c$ material constants and $\density$ density.  $\traceR \mathbf{A} = \sum_i A_{ii}$ denotes the flat space trace operator. Introducing pressure $\Pressure$ as adjoint force compensating volume work, the total stress $\stress = \stress^E + \stress^A$ can be separated into elastic and active contributions $\stress^E = \viscosity \DeformationTensor - \mathbb{I}\Pressure - \lambda \MolecularField + \Qten\MolecularField - \MolecularField \Qten$ and  $\stress^A = \activity \Qten$, respectively. Thereby, $\mathbb{I}$ is the identity, $\activity$ an activity parameter and $\viscosity$ denotes the shear viscosity. The terms with $\viscosity$ and $\Gamma^{-1}$  provide the dissipation mechanisms of the model. Following \cite{giomi2015geometry} we have neglected equilibirum/Ericksen stress as a second order effect. If $\activity = 0$ the model relates to the Beris-Edwards model \cite{beris1994thermodynamics}.

To derive a surface model we confine the equations to a thin film geometry $\surf_h = \surf \times [-h/2, h/2]$ with surface $\surf$ and constant thickness $h$ and formally let $h \rightarrow 0$. As discussed in previous works \cite{Nitschke_PRSA_2018,nitschke2019hydrodynamic} the choice of boundary conditions on $\partial \surf_h$ has a strong effect on the resulting thin film limit. On the surfaces $\surf$ and $\partial \surf_h$ we denote the outward oriented, \wrt\ $\surf$, surface normals by $\normal$ and $\normal_h$ which coincide for $h \rightarrow 0$. We further denote $\proj$ and $\proj_h$ as the associated surface projections, applied to each component. For the velocity we impose Dirichlet  boundary conditions on $\partial \surf_h$, $\Velocity\cdot \normal_h = 0$ and require normal stress condition for the deviatoric part $\proj_h [\DeformationTensor \cdot \normal_h] = 0$ and the rotational part $\proj_h [\VorticityTensor \cdot \normal_h] = 0$ of the rate of deformation tensor. Under these conditions the thin film limit of mass and momentum balance lead to results for the tangential velocity $\proj_h[\Velocity] \rightarrow \proj[\Velocity |_{\surf}] = \tangentVelocity$, see \cite{Miura_QAM_2018}. For the nematic ordering $\Qten$ we apply a tangential anchoring condition as detailed in \cite{Golovatyetal_JNS_2017,Nitschke_PRSA_2018}. We therefore split $\Qten$ in a tangential part $\proj_h[\Qten] = \qten_h$ and a normal part $\beta$ \wrt $\;$ to $\normal_h$ such that $\Qten = \qten_h - \beta/2\metric_h + \beta \normal_h\otimes\normal_h$ on $\partial\surf_h$, where $\metric_h$ denotes the associated surface metric. We impose Neumann boundary conditions $\proj_h[ \nabla \Qten \cdot \normal_h ] = 0$  on $\surf_h$. Analogue to velocity, we define the surface order tensor by the thin film limit $\proj_h[\Qten] \rightarrow \proj[\Qten |_{\surf}] = \qten$. Given the discussion of \cite{Nestler_SM_2020} we highlight the impact of the choice of $\beta$. With $\beta = 0$ we yield a surface model for uniaxial nematics \cite{Kralj_SM_2011,Jeseneketal_SM_2015}, which essentially has the same properties as a two-dimensional model, while $\beta \neq 0$ results in thin film models of nematics \cite{Napolietal_PRE_2012,Golovatyetal_JNS_2015,Golovatyetal_JNS_2017,Novack_SIAMJMA_2018,Nitschke_PRSA_2018,Nitschke_PRSA_2020} with several properties shared with three-dimensional models. The differences result from the consideration of extrinsic curvature contributions. Since the resulting models exhibit significant differences in the coupling mechanisms to geometry we will consider both conditions, $\beta = 0$ and $\beta = -S^*/3$, where $S^*= \sqrt{3/2}\| \Qten^*\|$ is defined as preferred degree of nematic ordering prescribed by the values of $a,\,b,\,c$. For the more details of the thin film limit we refer to \ref{sec:AppTFL}.

The thin film limit of the free energy results in $1 / h\;  \energy(\Velocity, \Qten) \rightarrow \energy_{\surf}(\tangentVelocity, \qten)$ with the surface free energy
\begin{align}
   \energy_{\surf}(\tangentVelocity, \qten) &= \int_{\surf} \frac{\density \tangentVelocity^2}{2} \,\mathrm{d}\surf + \frac{L}{2} \int_{\surf} \left\| \GradSurf\qten \right\|^2 + \left\| \shop \right\|^2\Tr\qten^2 -6\beta\meanc\left\langle \shop, \qten \right\rangle + C_1\,\mathrm{d}\surf \nonumber \\
& \quad+ \int_{\surf} \left(a -b \beta + \frac{3}{2} c \beta^2 \right) \Tr\qten^2 + c\Tr\qten^4 + C_2 \,\mathrm{d}\surf
\label{eq:surfaceenergy}
\end{align}
where $C_1 = C_1(\beta,\shop)$ and $C_2 = C_2(a,b,c,\beta)$ are constants. A set of covariant differential operators is used: $\GradSurf$ the covariant derivative, $\DivSurf$ the surface divergence, $\laplaceBochner = \DivSurf \GradSurf$ the Bochner Laplacian and $\levicivita{\tangentVelocity}{.}$ the directional derivative \wrt $\:$ $\tangentVelocity$. $\Tr \mathbf{a} = \sum_{ij} \mathbf{a}_{ij}\metric^{ij}$ denotes the trace operator in the curved space. We further use the curvature quantities: $\gaussc$ the Gaussian curvature, $\meanc$ the mean curvature and $\shop = - \proj[\nabla \normal]$ the Weingarten mapping of $\surf$. $\langle . \; ,\; .\rangle$ denotes the scalar product \wrt\ to the surface metric $\metric$. The molecular field is defined as $\molecularField = - \delta \energy_{\surf} / \delta \qten$, by variation \wrt\ trace free and symmetric tensors, which conforms to $\molecularField = \proj[\MolecularField  |_{\surf}]$, see \cite{Nitschke_PRSA_2018}. Finally by denoting $\pressure$ the surface pressure and labeling  $\deformationTensor = \proj[\DeformationTensor |_{\surf}]$ and $\vorticityTensor = \proj[\VorticityTensor |_{\surf}]$ the tangential parts of the symmetric and anti-symmetric rate of deformation tensor we yield the surface model
\begin{align}
   \partial_t\tangentVelocity + \levicivita{\tangentVelocity}{\tangentVelocity} - \viscosity\left(\laplaceBochner\tangentVelocity + \gaussc\tangentVelocity\right) + \GradSurf\pressure   + \friction \tangentVelocity &=  \lambda \DivSurf\molecularField + \DivSurf\left( \qten\molecularField - \molecularField \qten \right) \nonumber \\
   & \quad + \activity \DivSurf \qten  \label{eq:thinFilmLimitGiomiMomentum} \\
\DivSurf\tangentVelocity &=0 \label{eq:thinFilmLimitGiomiMass} \\
\partial_t\qten + \levicivita{\tangentVelocity}{\qten} + \vorticityTensor \qten - \qten \vorticityTensor &= - \lambda \deformationTensor + \Gamma^{-1}\molecularField \label{eq:thinFilmLimitGiomi_q}, 
\end{align}
where, following \cite{Pearce_NJP_2020}, an additional friction term $\friction \tangentVelocity$ is considered. With 
\begin{align} 
\molecularField &= L \laplaceBochner \qten - L (\meanc^2 - 2 \gaussc) \qten + 3 L \beta \meanc \left(\shop - \frac{1}{2} \meanc \metric \right) - \left(2a  -2 b \beta + 3 c \beta^2 \right) \qten \nonumber \\
& \quad - 2 c \Tr\qten^2 \qten \label{eq:molfield}
\end{align}
we observe the significant impact of the choice of $\beta$. For $\beta \neq 0$ an additional geometry coupling term emerges which forces the nematic director to align with minimal curvature lines. We will highlight the impact of the directed geometric contribution $f_{dir} = -6\beta\meanc \langle \shop, \qten \rangle$ in eq. \eqref{eq:surfaceenergy} and the resulting terms in the evolution equations in subsequent numerical experiments.

\section{Numerical approach}
\label{sec:2}
To solve the surface active nematodynamic model, eqs. \eqref{eq:thinFilmLimitGiomiMomentum}-\eqref{eq:molfield}, we us an implicit Euler scheme for temporal discretization. For each time step 
\begin{equation*}
    t^i \; \rightarrow \; t^{i+1}=t^i + \tau \; : \;  [\tangentVelocity^i,\, \pressure^i,\, \qten^i,\, \molecularField^i ] \; \rightarrow \; [\tangentVelocity^{i+1},\, \pressure^{i+1},\, \qten^{i+1},\, \molecularField^{i+1} ]
\end{equation*}
we apply an operator splitting approach. 
Therefore the surface Navier-Stokes-like equations are solved in a first step with $\qten^i,\, \molecularField^i$ of the previous time step. The equations are solved along the Chorin projection method \cite{Chorin_MC_1968} for a linearized transport term. As prediction step the intermediate velocity $\tangentVelocity^*$ is determined by 
\begin{align}
 \frac{1}{\tau}\tangentVelocity^{*} \!+\! \levicivita{\tangentVelocity^*}{\tangentVelocity^i} \!+\! \levicivita{\tangentVelocity^i}{\tangentVelocity^*} \!- \viscosity\left(\laplaceBochner\tangentVelocity^{*} \!+ \gaussc\tangentVelocity^{*}\right) \!&=\! \frac{1}{\tau}\tangentVelocity^{i} \!+ \lambda \DivSurf\molecularField^i \!+ \DivSurf\left( \qten^i\molecularField^i
 - \molecularField^i \qten^i \right) \nonumber \\
 &\quad + \activity \DivSurf \qten^i.  
\end{align}
This equation is solved by the component wise surface finite element method (SFEM) for vector-valued fields \cite{nestler2019finite}. For this purpose we use linear Lagrange elements on a surface triangulation. Surface normals and curvature quantities are given by analytical results on the nodes of the triangulation. If these quantities are not available we refer to \cite{Nitschkeetal_JFM_2012,Reuther_JFM_2020} for appropriate ways, how they can be computed. In the corrector step we evaluate $\pressure^{i+1}$ via a relaxation scheme $l\, \rightarrow \, l+1$
\begin{align}
 \frac{1}{\theta}\pressure_{l+1} + \tau \laplaceBeltrami \pressure_{l+1} = \frac{1}{\theta}\pressure_{l} + \DivSurf \tangentVelocity^*
\end{align}
with Laplace-Beltrami operator $\laplaceBeltrami$. We consider a stepsize $\theta = 10$ and $\pressure_0 = \pressure^i$ and iterate until $\| \pressure_l - \pressure_{l+1}\|_{L^2} / \| \pressure_{l+1}\|_{L^2} < 10^{-3}$. Given the updated pressure $\pressure^{i+1}$ the velocity is projected into solenoidal space such that 
\begin{align}
\tangentVelocity^{i+1} = \tangentVelocity^* - \tau \GradSurf \pressure^{i+1}.
\end{align}
As solution procedure we use scalar SFEM \cite{dziuk2013finite} with linear Lagrange elements. 

As second step of the operator splitting approach the Landau-de Gennes-like model is solved for updated $\tangentVelocity^{i+1}$ and it's associated fields $\deformationTensor
^{i+1},\, \vorticityTensor^{i+1}$. 
To evaluate the time step of the nematic quantities $\qten^{i+1}$ and $\molecularField^{i+1}$ we insert the molecular field, see eq. \eqref{eq:molfield}, into eq. \eqref{eq:thinFilmLimitGiomi_q}. Due to the strong nonlinearity for $c \gg L$ the resulting equation is solved via Newton iteration $l \, \rightarrow \, l+1$ with $\qten_0 = \qten^i$ until $\| \qten_l - \qten_{l+1}\|_{L^2} < 10^{-8}$. The Newton iteration step reads
\begin{align}
   & \frac{1}{\tau}\qten_{l+1} + \levicivita{\tangentVelocity^{i+1}}{\qten_{l+1}} + \vorticityTensor^{i+1} \qten_{l+1} - \qten_{l+1} \vorticityTensor^{i+1} + \lambda \deformationTensor^{i+1} \nonumber \\
   = & \frac{1}{\tau}\qten^{i} + \Gamma^{-1} \left[  \laplaceBochner \qten_{l+1} - L (\meanc^2 - 2 \gaussc) \qten_{l+1} - L S \meanc \left(\shop - \frac{1}{2} \meanc \metric \right)\right] \\
   & - \Gamma^{-1} \left[ \left(2a -2 b \beta + 3 c \beta^2 \right) \qten_{l+1} + 2 c \left[ \Tr\qten^2_{l} \,  \qten_{l+1} + 2 \langle \qten_l , \qten_{l+1} \rangle\, \qten_l   -2 \Tr\qten^2_{l} \, \qten_l \right] \right]. \nonumber
\end{align}
The system is solved by the component wise SFEM for Q-tensor-valued fields \cite{nestler2019finite} with linear Lagrange elements. Finally $\molecularField^{n+1}$ is evaluated by numerically solving the variational form of eq. \eqref{eq:molfield}.

The resulting linear systems are assembled and solved in the finite element toolbox AMDiS \cite{Veyetal_CVS_2007, Witkowskietal_ACM_2015} using a BiCGStab(l) solver. In Section \ref{sec:3} we discretize the surface by a surface triangulation with $2500$ vertices. Normals and curvature quantities are given analytically for the considered ellipsoidal shapes and are evaluated in the vertices. As time step size we use throughout the numeric experiments $\tau = 0.01$.

Applying component wise SFEM for vector- and Q-tensor-valued equations along \cite{nestler2019finite} we use the following penalty terms for extended quantities in normal direction
\begin{align}
    P(\underline{\tangentVelocity}, \boldsymbol \psi) &= \omega_n \int_{\surf} \normalC_i \normalC_j \psi_j \underline{\tangentVelocityC}_i  \,\mathrm{d}\surf, \\ 
    P(\underline{\qten}, \boldsymbol \Psi) &= \omega_n \int_{\surf} (\normalC_i \underline{\qtenC}_{ij})(\normalC_l \Psi_{lj}) + (\normalC_i \normalC_j \underline{\qtenC}_{ij})(\normalC_i \normalC_j \Psi_{ij}) \,\mathrm{d}\surf
\end{align}
with penalty parameter $\omega_n = 1000$ and appropriate test functions $\boldsymbol{\psi}$ and $\boldsymbol{\Psi}$. 

\begin{figure}
	\centering
	\includegraphics[width = 0.45\textwidth]{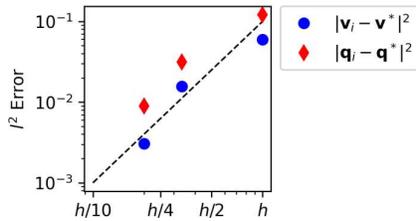}
	\caption{$L^2$ error convergence of solution procedure: $L^2$ errors for solution components $\tangentVelocity$ and $\qten$ \wrt\ solution evaluated on triangulation with $130K$ vertices. Black dashed line indicates quadratic convergence \wrt\ mesh size $h$.}
	\label{fig:numericValidation}
\end{figure}

For validation purposes a numerical convergence study is shown in Figure \ref{fig:numericValidation}. It considers the dynamics for $\activity = -8$ on a unit sphere for $t \in [75,75.25]$ on surface triangulations with $\left\{ 2.5k,\; 10k,\; 40k,\; 130k \right\}$ vertices and adjusted time steps $\left\{ 0.01,\; 0.005,\; 0.001,\; 0.0002 \right\}$. Using the solution on the triangulation with $130k$ vertices as reference we obtain quadratic error convergence in the solution components $\tangentVelocity$ and $\qten$.
This is in agreement with the obtained convergence results for the individual problems, the surface Navier-Stokes equations \cite{Reuther_PF_2018,Reuther_JFM_2020} and the surface Landau-de Gennes model \cite{Nitschke_PRSA_2018,Nitschke_PRSA_2020} and the expected results for the coupled problem in flat space.

\section{Results/Numerical Experiments}
\label{sec:3}
Various studies indicate a strong dependency of the position of nematic defects on curvature \cite{Kralj_SM_2011,bowick2009two,alaimo2017curvature}. While the stability of such configurations is well understood in passive systems, see e.g. \cite{Nestleretal_JNS_2017}, and the fundamental differences of the two systems, $\beta = 0$ and $\beta = -\orderp^*/3$, already explored \cite{Nitschke_PRSA_2018,Nitschke_PRSA_2020}, these studies neglect hydrodynamic effects. We here investigate using numerical experiments how curvature does impact the dynamics of active systems with hydrodynamics. Observed dynamics range from stationary to periodic and turbulent regimes where activity is the relevant parameter, see \cite{keber2014topology} for experimental observations. To limit the complexity in the numeric investigations we focus on activity parameter values close to the transition between the stationary and the periodic regime. Furthermore we will consider 
'small' geometries with strong curvatures typical for microscopic biological active systems \cite{keber2014topology}. For this purpose, we non-dimensionalise the system along a characteristic length of domain $l^* = 1$, velocity $\tangentVelocity^* = 1$ and, for sake of simplicity, we consider the parameters $\viscosity = 1$, $\Gamma^{-1}L = 1$ such that the Ericksen number is fixed at $\EricksenNumber = \viscosity l^* \tangentVelocity^*/ L \Gamma^{-1}= 1$ while the Reynolds number $\ReynoldsNumber = \density l^* \tangentVelocity^* / \viscosity$ scales by $\density$. If not noted otherwise we use $\density = 1$. Furthermore, we chose the alignment parameter $\lambda = 0.7$, as in \cite{thampi2014instabilities}, the friction coefficient $\friction=0.05$ and the parameters in the Landau-de Gennes energy as $a=-10$, $b=0$ and $c=5$ to maintain stable nematic texture and defect core covering of approximately $7\%$ of the surface. We focus on extensile activity, with $\activity \in [-10, 0]$, such that the dynamics is far from transition to low Reynolds number turbulence. For studies in this regime we refer to \cite{Pearce_PRL_2019,Mickelin_PRL_2018,Rank_PF_2021} and for other related approaches to \cite{Napoli_PRE_2020,Pearce_NJP_2020}.

\begin{figure}
	\centering
	\includegraphics[width = 0.99\textwidth]{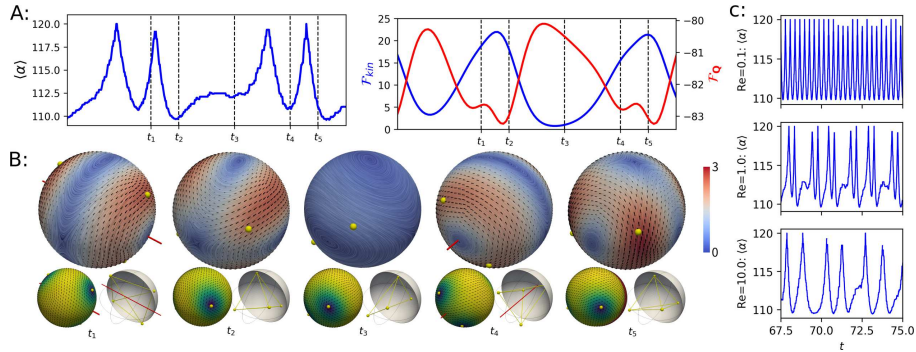}
	\caption{\textbf{Impact of Flow Regime for Active Nematics on Sphere $\activity=-8$:} [A] Details of single period of dynamics at $\Re\,=\,1$. (left) average angle of four defect configuration indicating oscillation between tetrahedral $\langle \alpha \rangle =120$ and planar $\langle \alpha \rangle =110$ configuration via pairwise rotation along an axis. (right) Kinetic and nematic energy for period of dynamics. [B] Snapshots of flow (top line, black arrow - direction, color - magnitude), and nematics (left: texture and $\|\qten\|$ colors, right: schematic of defect configuration). Red lines indicate axis of defect pair rotation. Snapshots from left to right correspond to $t_1,\,t_2,\,t_3,\,t_4,\,t_5$ marked in [A] such that $t_1$ for pairwise rotation at axis near planar defect configuration, $t_2$ breakdown of defect pairs in tetrahedral configuration, $t_3$ stagnation of flow, $t_4$ rearranged pair rotation and $t_5$ again breakdown of pairs. [C] Long term behavior of defect configuration depending on $\Re$. (top) For $\Re=0.1$ we observe single frequency in oscillations. (mid) At $\Re = 1$ inertia effects become traceable, slowing down axis rearrangement and inducing (periodic) stagnation events. (bottom) $\Re=10$ yields aperiodic configurations due to complex interaction of elastic stresses and inertia.}
	\label{fig:sphereExperiments}
\end{figure}

We consider a sphere, and prolate and oblate geometries. The first allows for comparison with the experimental results in \cite{keber2014topology} and detailed studies on the influence of $\ReynoldsNumber$ on flow, nematic texturing patterns and defect dynamics, see Figure \ref{fig:sphereExperiments}. The other geometries are used to study the influence of varying curvature. For the sphere the differences between the two systems $\beta = 0$ and $\beta = -\orderp^*/3$ are negligible. We therefore systematically only compare the two systems $\beta = 0$ and $\beta = -\orderp^*/3$ for the ellipsoidal geometries. For all considered numerical experiments four $+ 1/2$ disclination are present and we analyse their dynamics. We apply principal component analysis (PCA) to the combined set of defect trajectories to determine the number of dominant temporal frequencies of the dynamic regime. Along this number we group the evaluated dynamics into three classes: stationary, single frequency periodicity and dynamics with increased complexity, for details see \ref{sec:methods}.
All numerical experiments are concerned with the behaviour of the late dynamics such that the impact of initial values can be neglected, for details on the setup see \ref{sec:initValue}.

\paragraph{Unit Sphere Experiments}

We evaluate the dynamics for $\activity = -8$ in the time domain $[0,\, 75]$ for $\ReynoldsNumber = 0.1,\,1,\, 10 $, see Figure \ref{fig:sphereExperiments}. To focus on the periodic dynamics we only analyse the flow patterns and nematic textures in the late time domain $t \in [70,\, 75]$. We observe oscillations between planar and tetrahedral defect configurations, which we quantify using the average defect angle 
\begin{equation*}
    \langle \alpha \rangle = \frac{1}{6} \sum_{i=1}^{4} \sum_{j>i}^{4} \cos(\xb^i_d \cdot \xb^j_d), 
\end{equation*} 
with $\xb_d^i$ the defect positions. 
%A more detailed view highlights the importance of hydrodynamic interactions. The kinetic and nematic energy for a period of dynamics indicate that through pairwise rotation nematic ordering relaxes to an alignment with flow direction reducing elastic stresses and increased kinetic energy. After a second rotation a critical velocity is reached and defects break rotation and rearrange to new pairs along new axis. Inertia of flow induces every second rearrangement an excessive displacement of defects yielding strong elastic stresses and stagnation of flow, see Figure \ref{fig:sphereExperiments}[A](right). These details have not been analysed in \cite{keber2014topology}, but in principle 
In the case $\Re=0.1$ (Stokesian like regime) the experimental observations of \cite{keber2014topology} are reproduced and a single frequency oscillation is observed. For $\Re=1.0$ we yield an additional frequency in the periodic oscillations and no periodicity is found for $\Re=10$. We conclude that increasing inertia effects play a substantial role on the destabilization of the oscillating defect configurations. However, spontaneous nematic distortion, as described in \cite{thampi2014instabilities, giomi2015geometry}, is not observed in the chosen parameter regime.

\begin{figure}
	\centering
	\includegraphics[width = 0.99\textwidth]{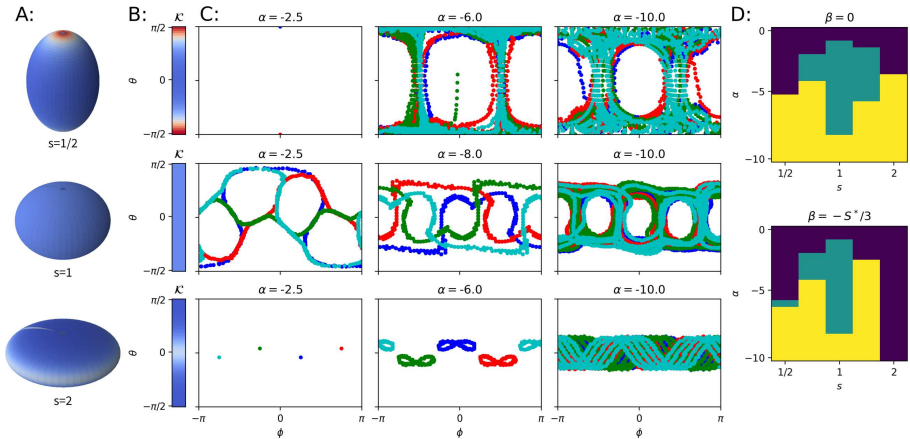}
	\caption{\textbf{Impact of Curvature Modulations on Defect Trajectories:} [A] Considered ellipsoidal geometries of equal surface area. (top) Strong prolate geometry $\deformParam=1/2$, (mid) unit sphere $\deformParam=1$ and (bottom) strong oblate geometry $\deformParam=2$. Colors encode Gaussian curvature $\gaussc \in [6.89,\, 0.17]$. [B] Distribution of Gaussian curvature along azimuthal angle $\theta$. [C] Typical defect trajectories (described by spherical coordinates $(\phi,\theta)$, each color denotes one defect) of active nematodynamics for various activities (columns) and geometries (lines). [D] Classification of defect dynamics for activity $\activity$ and geometry $\deformParam$ variation for $\beta = 0$ and $\beta=-S^*/3$. Dark blue indicates stationary defects, cyan stands for single frequency periodic trajectories and yellow indicate trajectories with complex periodicity.}
	\label{fig:DefectTrajectoriesSphereoidsBetaZero}
\end{figure}

Given these results we vary $\activity \in [-10,\, 0]$ and explore the impact of activity on the defect patterns, see Figure \ref{fig:DefectTrajectoriesSphereoidsBetaZero}(mid line). The defect trajectories are represented in spherical coordinates for three different activity values. In general we observe four intertwined defect trajectories oscillating between regular tetrahedral and planar configurations. In the light of investigations for a simplified defect particle model in \cite{brown2020theoretical}, we asses the ratio of defect speed and rearrangement time of nematic texture as crucial to enable two different motion patters. For low activity, with low defect speed, the nematic texture can relax sufficiently fast and elastic stresses are dominated by active stresses. Therefore the defects move in more or less straight lines (geodesic circles). Increasing the activity and defect speed crosses eventually a threshold where elastic stresses no longer quickly relax and periodicity dominates the active stresses. In such situations we observe a breakdown of flow and defect motion. In this stagnation phase the nematic texture relaxes and the flow rearranges, see Figure \ref{fig:sphereExperiments}[A,B]. After this rearrangement the defect motion restarts but with a significant change in direction, see the kinks in defect trajectories in Figure \ref{fig:DefectTrajectoriesSphereoidsBetaZero}(mid line) at $\activity=-8$. Further increase of activity yields higher magnitudes in flow such that inertia effects break down the strict periodicity of previous trajectories while the overall motion structure (oscillation between tetrahedral and planar configurations) remains.

\paragraph{Interplay of Curvature and Defect Dynamics}
Deforming the unit sphere towards prolate and oblate ellipsoidal geometries with the same surface area breaks the symmetry. We characterize the geometries along a deformation parameter $\deformParam = A/C$ where $A$ and $C$ denote the ellipsoidal body axes in $X$ and $Z$ direction. The resulting geometries only have positive Gaussian curvature. Most significant features are areas of increased and reduced curvature. In the prolate case curvature is concentrated at the poles while in oblate geometries high curvature areas form a manifold at the rim of the geometry, see Figure \ref{fig:DefectTrajectoriesSphereoidsBetaZero}[A,B]. Investigations for passive nematics \cite{bowick2009two,Kralj_SM_2011,Nitschke_PRSA_2018,Nitschke_PRSA_2020} have established that defects are attracted to regions where curvature matches their topological charge. Therefore we expect additional geometric forces impacting the dynamics of defect motion. These effects result from explicit curvature terms and curvature sensitivity of the covariant derivative in $\molecularField$ and subsequent forces in momentum balance via $\DivSurf(\stress^E)$. Now, also the differences between the two systems $\beta = 0$ and $\beta=-S^*/3$ will become apparent. We evaluate the dynamics for a set of five deformations $\deformParam \in \left\{ 1/2,\; 3/4,\; 1,\; 5/4,\; 2 \right\}$ and 20 activity values $\activity = [-10,0]$. Figure \ref{fig:DefectTrajectoriesSphereoidsBetaZero}[C] shows typical examples of observed defect trajectories for the system with $\beta=0$. The corresponding results for $\beta=-S^*/3$ are qualitatively similar for the prolate and sphere geometries. The differences for the oblate geomatries will be discussed below.

In case of strong prolate geometry $\deformParam=1/2$, see Figure \ref{fig:DefectTrajectoriesSphereoidsBetaZero}[C](top line), the defects are located predominantly close to the poles. In case of weak activity the curvature forces dominate and fix the positions of defects and even suppress motion. The geometric forces even unify the two $+1/2$ defects on the poles to form $+1$ defects as already described in \cite{Kralj_SM_2011} for passive systems. The behaviour is maintained for $|\activity| < 5$. For activity beyond this threshold the geometric forces are outweighed and defect motion occurs. Here we observe a periodic motion consisting of two elements. The first is a circling motion of defect pairs around the poles and the second is a periodic breakdown of these pairs where on each pole a single defect leaves the pole region, crosses the low curvature region at the equator and aligns with the remaining defect at the opposing pole. With increasing activity, as in the unit sphere case, higher velocities emerge, which eventually destabilizes the defect trajectories, by inertia effect.

An example for oblate geometries is shown in Figure \ref{fig:DefectTrajectoriesSphereoidsBetaZero}[C](bottom line) for $\deformParam=2$. Again for low activity $|\activity| < 3.5$ a stationary defect configuration emerges, where the four $+1/2$ defects are attracted to the rim in a quasi planar configuration. In this configuration the geodesic distance between the defects is no longer maximized. We thus conclude the localisation of the defects to be a geometric effect. Increasing activity yields a breakdown of this stationary configuration and the emergence of a periodic defect movement pattern. Remarkably, we observe a large variety of defect motion patterns drastically rearranging through a small change in activity magnitude. Based on ideas of \cite{zhang2020dynamics} for the spherical case, we suspect these patterns to emerge from the relation of defect motion wavelengths and characteristic curvatures and lengths in the geometry. To classify the emerging variety of patterns is beyond the scope of this paper and has to be investigated elsewhere. As in the spherical and prolate case we yield for higher activity a destabilization of movement patterns.

We conclude these numerical experiments by a systematic evaluation and classification of defect trajectories. The results are summarized in the form of a phase diagram in Figure \ref{fig:DefectTrajectoriesSphereoidsBetaZero}[D](top). For sake of clarity the observed dynamics are clustered along three groups: (i) stationary, (ii) single frequency and (iii) complex dynamics. For details regarding this classification see \ref{sec:methods}. We observe two prominent effects. First, breaking the symmetry of geometry introduces additional geometric forces acting on nematic texture, prescribing a preferred defect position. Compared with the unit sphere case these stationary defect configurations persist for stronger activities for prolate and oblate geometries. As second effect we highlight the impact on trajectory periodicity. Even mild geometry deformations yields a strong reduction of activities enabling single periodic movement patterns. In the case of strong geometry deformations the range of single periodic movement is below our activity testing resolution such that we observe an immediate transition from stationary to complex trajectories. Contrary to the first effect, this effect is stronger for prolate geometries compared to oblates. Recalling the distribution of Gaussian curvature, see Figure \ref{fig:DefectTrajectoriesSphereoidsBetaZero}[B], this can be attributed to the different magnitudes in curvature of the deformations, \eg\ in strong prolate poles $\gaussc \approx 6.5$ and strong oblate rim $\gaussc \approx 2.5$. 

\begin{figure}
	\centering
	\includegraphics[width = 0.99\textwidth]{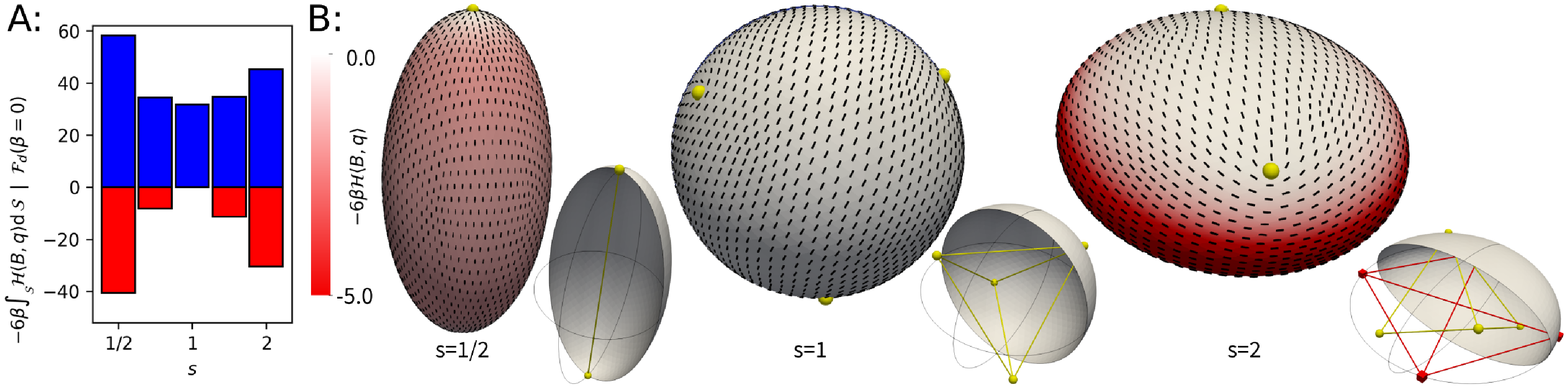}
	\caption{\textbf{Directed Geometric Forces in System with $\beta=-\orderp^*/3$:} [A] Distortion energy of steady state for geometries $\deformParam \in [1/2, 2]$ for $\beta =0 $ (blue bars) vs directed geometric contribution by $f_{dir} = -6\beta\meanc \langle \shop, \qten \rangle$ (red bars) for $\beta = -\orderp^*/3$. [B] Curvature distribution of geometry impacts strength of directed geometric energy (color code) and final nematic texture (black lines). At $\deformParam=1/2$ curvature is concentrate on poles where at each pole two $+1/2$ merge to one $+1$ defect, see insert schematics. Defects with $\|\qten\|=0$ cancel directed geometric contribution at poles while nematic texture aligns with lines of minimal curvature on other parts of geometry with limited directed geometric contribution $f_{dir} \approx -1.5$. For the unit sphere $\deformParam=1$ geometry provides no prefered alignment $f_{dir} = 0$ such that typical tetrahedral configuration (insert schematics) is observed for $\beta=0$ and $\beta=-\orderp^*/3$. At strong oblate geometry $\deformParam=2$ the curvature is focused at the rim manifold yielding there strong directed geometric energy contribution $f_{dir} < -5$. The nematic texture aligns parallel with the rim, pushing defects off the rim and forming a tetrahedral configuration, see yellow configuration in insert schematic. This is not observed for $\beta=0$ where defects align with the rim in an planar configuration, see red configuration in insert schematics.}
	\label{fig:GeometricForce_Sphereoids}
\end{figure}

Figure \ref{fig:DefectTrajectoriesSphereoidsBetaZero}[D](bottom) shows the corresponding phase diagram for the system $\beta = -S^*/3$. The comparison of the phase diagrams highlights the quantitative differences between the two systems $\beta = 0$ and $\beta = -S^*/3$. These differences can be explained by two effects. The first is present already in the passive case, $\activity=0$, and does not required hydrodynamic interactions. The additional directed geometric forces in the system with $\beta = -S^*/3$ have an impact on the nematic texture \cite{Nestler_SM_2020}. Reviewing the contributions of isotropic and directed distortion energy in the steady state configurations, see Figure \ref{fig:GeometricForce_Sphereoids}[A], we observe a strong directed contribution providing a strong energy sink for geometries with broken symmetry. These contributions reach a magnitude of approximately half of the isotropic energy for strong deformations. Turning to the spatial distribution of the directed distortion energy density, see Figure \ref{fig:GeometricForce_Sphereoids}[B], shows low magnitudes covering the bulk of prolate geometries, while on oblate geometries the energy density reaches high magnitudes on the rim with almost zero magnitude on quasi planar top and bottom part. As detailed in \cite{Nestler_SM_2020} this distribution can be attributed to the alignment of nematic texture with lines of minimal curvature. The stationary defect configurations for pure isotropic distortion ($\beta=0$) and including directed distortion energy ($\beta = -\orderp^*/3$) match for prolate geometries and the sphere. But for oblate geometries, with its high localized directed geometric energies and subsequent strong forces, we yield an almost in plane defect configuration for $\beta = 0$, while for $\beta = -\orderp^*/3$ the defects form a distinct tetrahedral configuration.

\begin{figure}
	\centering
	\includegraphics[width = 0.99\textwidth]{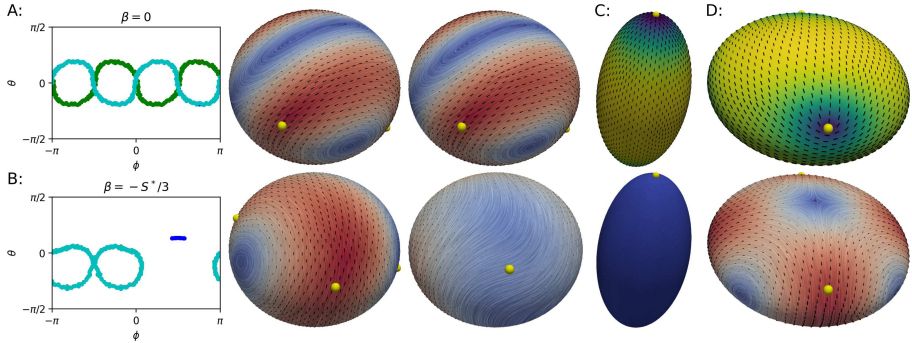}
	\caption{\textbf{Impact of Directed Geometric Forces on Defect Dynamics:} [A] Defect trajectories(left)  on mild oblate deformation $\deformParam = 5/4$ at $\activity = -3.5$ for $\beta=0$ exhibit pairwise defect rotation similar to Fig \ref{fig:DefectTrajectoriesSphereoidsBetaZero}. Here trajectories cross the high curvature rim but defect pair rotation axes are fixed to the plane of the rim. Flows(mid: front, right: back, black arrows indicate direction, colors for magnitude) consist of two counter rotating vortices. [B] Same setup for $\beta = -S^*/3$ yields dynamics with broken symmetry such that three defects align to two frequency configuration below oblate rim and a single quasi stationary defect above the rim. A strong central jet(mid: front) in double vortex structure at three defect configuration and weak flow at isolated defect(right: back). [C] For strong prolate deformation $\deformParam=1/2$ at $\activity=-5$ and $\beta = -S^*/3$ a stationary nematic texture(top: black strokes, colors encode $\|\qten\|$) is observed, exhibiting rotational symmetry with two $+1$ defects(yellow dots) on poles yielding no flow(bottom). [D] On strong oblate geometry $\deformParam=2$, $\activity=-5$ and $\beta = -S^*/3$ nematic texture(top) exhibits broken symmetry with four $+1/2$ defects. Defects induce strong flows(bottom), but geometric forces on nematic texture dominate and no defect motion is observed.}
	\label{fig:DynType_Sphereoids}
\end{figure}

Considering the active dynamics as a destabilization of the passive nematic texture and defect configuration explains most of the differences in the phase diagrams for the system with $\beta = 0$ and $\beta=-\orderp^*/3$ in Figure \ref{fig:DefectTrajectoriesSphereoidsBetaZero}[D]. Already for weak distortions of the sphere in prolate and oblate direction $\deformParam \in \left\{ 3/4,\; 5/4 \right\}$ a distinct impact of directed geometric forces emerge. Curvature forces suppress dynamics for low activity such that the transition to the simple dynamics regime is shifted to stronger activity. The remarkable effect of the system with $\beta=-\orderp^*/3$ is the earlier onset of complex dynamics (compared to spherical case) effectively shrinking the domain of simple dynamics for increasing geometric distortions. This effect can be addressed to the situation that curvature drives the defects to configuration with non-maximal distance. As flow is associated with defects the resulting closer defect proximity yields stronger flows. Subsequently inertia effects are increased and 
'earlier' transition to complex defect dynamics occurs. Such chain of effects culminates for strong curvatures in the direct transition of stationary defects to complex defect dynamics. Especially in the case of strong prolate distortion, where on each pole two $+1/2$ defects merge to form one radial symmetric $+1$ defect with an associated stable radial symmetric nematic texture which does not induce any flow. See Figure \ref{fig:DynType_Sphereoids}[C]. This drastically changes as soon as the symmetry is broken and the described effect cascade yields strong flows and immediately complex defect dynamics. However, as in the passive case, the differences between the two systems $\beta = 0$ and $\beta=-\orderp^*/3$ are limited for prolate geometries. This changes for oblate geometries, where a decisive difference in the phase diagram is observed. Here the immediate transition between stationary to complex defect dynamics occurs already for weak oblate deformation, indicating a decisive impact of directed geometric forces on defect dynamics, see Fig \ref{fig:DynType_Sphereoids}[A,B]. 

While in the case of $\beta=0$, Fig \ref{fig:DynType_Sphereoids}[A], we observe a pairwise, single frequency, defect rotation similar to the unit sphere, the case of $\beta = -S^*/3$, Fig \ref{fig:DynType_Sphereoids}[B], exhibits two defect motion patterns. A single defect remains quasi stationary on the upper half of the oblate rim, while the remaining three defects form a two frequency moving pattern. This 'separation' of dynamics is facilitated by the directed geometric forces inducing the preferred nematic alignment along the rim such that defects tend to move parallel with the rim. Finally the directed forces become dominant for strong oblate deformation such that stationary defect dynamics are observed across the complete range of activity $\activity \in [-10,0]$. As discussed in the passive case the ringlike domain of high curvature of the oblate geometries induces a forcing on the nematic texture to align and thus is effectively limiting the possible defect positions. Yet these geometric forces fix the nematic texture, see Figure \ref{fig:DynType_Sphereoids}[D](top), but do not suppress flows. Contrary to the prolate case with a radial symmetric texture and two $+1$ defects, the oblate case yields a nematic texture with broken symmetry and subsequent strong flows, see Figure \ref{fig:DynType_Sphereoids}[D](bottom). Such phenomena of stationary defects and non zero flows could in principal also be present for $\beta = 0$ but only for much lower magnitudes of activity and thus for much weaker flows. The stationary defects and non zero flows contradict the assumption in coarse grained models of active nematodynamics, which consider defects as active particles, solely responsible for the induced flow \cite{giomi2014defect} and thus provide an other example of the limited applicability of this modeling approach on curved surfaces.

\section{Conclusion}
We derived a surface model for active nematodynamics by considering a thin film limit of a hydrodynamic liquid crystal theory. This ensures physical significance in the sense of thermodynamic consistency and independence of chosen coordinates of the model. By specifying the normal components of nematic ordering $\beta = \normal \Qten \normal$ on the boundary of the thin film we consider intrinsic and extrinsic curvature contributions in the derived surface model. We set up a numeric solution procedure by combining surface finite element approaches for vector- and tensor-valued surface PDEs \cite{nestler2019finite} in an operator-splitting approach and established in numerical experiments quadratic $L^2$ error convergence \wrt $\;$mesh refinement in tangential surface fluid velocity $\tangentVelocity$ and tangential surface Q-tensor $\qten$.

The numerical approach is used to explore the coupling of active dynamics and geometry in small domains with strong curvature. We validate the approach on a sphere and analyse the effect of hydrodynamics on the experimentally found oscillations between tetrahedral and planar defect arrangements \cite{keber2014topology}. Deforming the sphere to prolate and oblate geometries allows the curvature to shape defect trajectories, creating a wide range of movement patterns beyond these oscillations. We classify the defect dynamics into stationary defects, single frequency periodic trajectories and trajectories with complex periodicity. We found that the tight coupling of nematics/geometry and flow/geometry has a strong impact on the transitions between these regimes of defect motion. Especially if extrinsic curvature effects are taken into account, $\beta = -S^*/3$, the distribution of curvature has a decisive impact on dynamics. In the case of concentrated curvature in point like areas, as in strong prolate deformations, the coupling mechanisms act predominantly on defect position, while the bulk of nematic texture is not affected. In this case the regime of stationary defects extends to medium activity ranges and suppresses any flow. Quite contrary for shapes with extended areas of high curvature, as in the oblate geometry, the defects and nematic texture are affected by curvature yielding a geometric forcing which limits the range of defect trajectories. In this case situations for medium activities occur where defect configurations are stationary but strong flow patterns are present. Beyond an activity threshold, which strongly depends on the geometry, the geometric forces determine the defect motion and can be used to control them. We have only concentrated on the onset of these motion patterns. To fully understand the complex periodicity of the defect motion patterns would require further investigations. As started for spherical surfaces \cite{zhang2020dynamics} also for non-spherical surfaces the coupling of characteristic geometric lengths and defect movement pattern wavelength seems crucial to understand the mechanism of selection of motion patterns. The proposed model in principle provides the numerical tools to analyse this for general surfaces.

\appendix

\section{Thin Film Limit}
\label{sec:AppTFL}
To perform the thin film limit of the state equations \eqref{eq:stateEqu} in full detail includes extensive calculations as presented in \cite{Nitschke_PRSA_2018} for passive dry nematics or \cite{Miura_QAM_2018,nitschke2019hydrodynamic} for the Navier-Stokes equations. Therefore, we will present here only essential steps and refer for detail to \cite{Nitschke_PRSA_2018,nitschke2019hydrodynamic}. 

For velocity we impose homogeneous Dirichlet boundary condition $\Velocity \cdot \normal_h=0 \mbox{ on } \partial \surf_h$. By \cite{Nitschke_PRSA_2018}(Lemma 7), such conditions continue into the thin film such that
\begin{align}
    \Velocity \cdot \normal=\landau(h),\quad \partial_{\normal}(\Velocity \cdot \normal)=\landau(h^2),\quad \partial^2_{\normal\normal}(\Velocity \cdot \normal)=\landau(h^2)\quad \mbox{ on } \surf. \label{eq:AppVeloBC}
\end{align}
This immediately establishes the thin film limit of the incompressibility condition 
\begin{align}
\nabla \cdot \Velocity = 0  \underset{h \rightarrow 0}{\rightarrow } \DivSurf \tangentVelocity = 0. 
\end{align}
We also impose the normal stress boundary condition $\proj_h[\stress \cdot \normal_h] = 0 \mbox{ on } \partial \surf_h$ which can be interpreted as the right normal part of $\stress$ does not contain any tangential parts. In this sense we can write equivalently $\tb \cdot [\stress \cdot \normal_h] = 0\; \forall \tb \in T\surf_h$, where $T\surf_h$ denotes the tangent bundle of $\surf_h$. Such condition again continues into the thin film and yields
\begin{align}
    \tb \cdot [\stress \cdot \normal]=\landau(h),\quad \tb \cdot \partial_{\normal} [\stress \cdot \normal]=\landau(h^2),\quad \tb \cdot \partial^2_{\normal\normal} [\stress \cdot \normal]=\landau(h^2)
\end{align}
$\forall \tb \in T\surf$ on $\surf$. Given these conditions and defining the surface stress by $\stress_{\surf} = \proj[\stress |_{\surf}]$ yields
\begin{align}
    \nabla \cdot \stress |_{\surf} = \DivSurf \stress_{\surf} + \partial_{\normal} (\stress|_{\surf} \cdot \normal ) + \shop \cdot (\stress|_{\surf} \cdot \normal ) + \landau(h) \underset{h \rightarrow 0}{\rightarrow } \DivSurf \stress_{\surf}. \label{eq:AppThinfilmDiv}
\end{align}
Combined with the limit $h \rightarrow 0$ for $D \Velocity / D\,  t \rightarrow D \tangentVelocity / D t$ by \eqref{eq:AppVeloBC} the thin film limit of the tangential part of momentum balance follows. Inserting now the definition of total stress $\stress = \viscosity \DeformationTensor - \mathbb{I}\Pressure - \lambda \MolecularField + \Qten\MolecularField - \MolecularField \Qten + \activity \Qten$ we can check the conformity of the chosen boundary conditions for $\Velocity$ and $\Qten$ with the normal stress condition of the total stress. For isotropic stress $\mathbb{I}\Pressure$ the normal stress condition holds as well as for the deviatoric part $\viscosity \DeformationTensor$ subject to homogeneous Navier boundary conditions $\proj[ \DeformationTensor \cdot \normal_h ] = 0$. The normal anchoring condition for the $Q$ tensor $\normal_h \cdot \Qten\cdot \normal_h = \beta \mbox{ on } \partial \surf_h$ and the molecular field $\normal_h \cdot \MolecularField\cdot \normal_h = 0 \mbox{ on } \partial \surf_h$ also conform, via \cite{Nitschke_PRSA_2018}(Lemma 7), to the normal stress condition enabling the usage of the limit \eqref{eq:AppThinfilmDiv} in the momentum balance for active nematics.
Finally, by using normal anchoring of $\Qten$ and $\nabla_{\normal_h} \Qten = 0 \mbox{ on } \partial \surf_h$ we can establish the thin film limit $h \rightarrow 0 $ of the Laplacian
\begin{align}
\nabla \cdot \nabla \Qten |_{\surf}  \underset{h \rightarrow 0}{\rightarrow } \laplaceBochner \qten.
\end{align}
and yield, in combination with $\proj[\VorticityTensor \cdot \normal_h] = 0 \mbox{ on }\partial\surf_h$, the thin film limit of the nematic state equation.

\section{Data Analysis}
\label{sec:methods}
To obtain a qualitative understanding of the evolution of active nematodynamics we apply Principal Component Analysis (PCA), see \eg\ \cite{wold1987principal}, to the trajectories of the nematic defects. 

As first step we evaluate the dynamics of eqs. \eqref{eq:thinFilmLimitGiomiMomentum}-\eqref{eq:thinFilmLimitGiomi_q} in spatial and temporal resolution via the described SFEM on a time domain $t \in [0,500]$. For the nematic texture we evaluate $\|\qten\|$ and use regions with $\|\qten\| < S^*/4 $ to identify defect positions within each time step. In temporal direction we assemble these positions to trajectories $\trajectory_n(t)$, $n \in \{ 1 \hdots N\}$ by using the particle tracking capacities of MOSAIC toolbox\cite{sbalzarini2005feature}. From the defect trajectories at time $t_i$ we assemble the defect positions $\trajectory_n(t_i)$ into a configuration vector $C(t_i) = [\trajectory_1(t_i), \hdots, \trajectory_N(t_i)]^T$. To suppress the impact of initial conditions we include only results of late dynamics. Given $M$ time steps $t_i \in [450,\, 500]$ we yield the data set to apply PCA. 

In short, we assemble the configuration snapshots of $N$ trajectories to the data matrix $X \in \mathbb{R}^{M\times 3N}$. With the zero mean data $B = X - [1,\hdots ,1]^T \otimes [  \sum_{i=1}^{3N} \sum_{j=1}^{M}X_{ij}\mathbf{e}_j] /M $, we obtain the covariance matrix $C=1/(M-1) B^T \cdot B \in \mathbb{R}^{3N \times 3N}$ and its eigenvalue $\lambda^i$ and eigenvector $V^i$ decomposition. Assuming a decreasing order of eigenvalues, each $\lambda^i$ encodes the variance captured in the direction $V^i$. Therefore, we select a subset of eigenvalues $i \in \{1,\hdots,L\}$ such that $99\%$ of the covariance is captured. The associated eigenvectors are compiled to the matrix $W = [V^0, \hdots, V^L] \in \mathbb{R}^{3N \times L}$ which is used to project the data onto this eigenvector basis $A = B \cdot W$. By this definition the $i$-th column of $A$ contains the PCA coefficient function associated with $\lambda^i$ and  $V^i$.

For the purpose of regime classification we use the average trajectory length, the number of eigenvalues necessary to capture $99\%$ variance and an estimate of periodicity in the PCA coefficients function. A stationary defect dynamic is given if the total covariance $\sum_i \lambda_i < 10^{-4}$ across time domain $[450, 500]$ vanishes. For instationary cases we classify the dynamics as single frequency if $L\leq3$ and all associated PCA coefficient functions can be approximated by a periodic expansion by a sample of the coefficient function within a $20\%$ relative $l^2$ error bound. Hereby, the expansion sample length is given by the global peak(beside identity) of the autocorrelation of the PCA coefficient function. Any other dynamics we classify as complex or with increased complexity.

\section{Initial Values}
\label{sec:initValue}
To obtain an inital four $+1/2$ defect configuration we start from a nematic texture consisting of two $+1$ vortex defects and evaluate its evolution for $t \in [0,0.05]$ in a dry $\tangentVelocity \equiv 0$ and passive $\activity = 0$ setup. By this preliminary evolution we yield a smooth nematic texture, where defects are distributed slightly unbalanced along the XY equator of the spheroid and exhibit $\|\qten\| = 0$ in core regions.

The inital two defect nematic texture is evaluated by splitting the geometry $\surf$ by the XZ plane. On each half a constant vector field $\mathbf{P}$ is projected to $T\surf$ and normalized. With $\mathbf{p}_0 = \proj[\mathbf{P}]/\|\proj[\mathbf{P}]\|  \times \normal $ we define the nematic texture $\qten^0 = \orderp^* ( \mathbf{p}_0 \mathbf{p}_0 - 1/2 \metric )$. For $y>0:\; \mathbf{P} = [0,1,0]$ and $y\leq 0:\; \mathbf{P} = [0.1,1,0.1]$.

%\begin{thebibliography}{99}
%\bibitem{Berger}M. J. Berger and P. Collela, Local adaptive mesh refinement
%for shock hydrodynamics,
%J. Comput. Phys., 82 (1989), 62-84.
%\bibitem{deBoor}C. de Boor,  Good Approximation By Splines With Variable Knots II, in Springer Lecture
% Notes Series 363, Springer-Verlag, Berlin, 1973.
%\bibitem{TanTZ} Z. J. Tan, T. Tang and Z. R. Zhang, A simple moving mesh method for one- and
%two-dimensional phase-field equations, J. Comput. Appl. Math., to appear.
%\bibitem{Toro}E. F. Toro, Riemann Solvers and Numerical Methods for Fluid Dynamics,
%Springer-Verlag Berlin Heidelbert, 1999.
%\end{thebibliography}

%%%% Acknowledgments %%%%%%%%
\section*{Acknowledgments}
We acknowledge financial support by DFG through FOR3013 and computing resources provided by PFAMDIS at FZ J\"ulich.

%%%% Bibliography  %%%%%%%%%%
\bibliographystyle{abbrv}
\bibliography{lib}

\begin{thebibliography}{10}

\bibitem{alaimo2017curvature}
F.~Alaimo, C.~K{\"o}hler, and A.~Voigt.
\newblock Curvature controlled defect dynamics in topological active nematics.
\newblock {\em Sci. Rep.}, 7:1--9, 2017.

\bibitem{Apaza_SM_2018}
L.~Apaza and M.~Sandoval.
\newblock {Active matter on Riemannian manifolds}.
\newblock {\em {Soft Matter}}, {14}:{9928--9936}, {2018}.

\bibitem{beris1994thermodynamics}
A.~N. Beris and B.~J. Edwards.
\newblock {\em Thermodynamics of flowing systems with internal microstructure}.
\newblock Oxford University Press, 1994.

\bibitem{bowick2009two}
M.~J. Bowick and L.~Giomi.
\newblock Two-dimensional matter: order, curvature and defects.
\newblock {\em Adv. Phys.}, 58(5):449--563, 2009.

\bibitem{brown2020theoretical}
A.~T. Brown.
\newblock A theoretical phase diagram for an active nematic on a spherical
  surface.
\newblock {\em Soft Matter}, 16(19):4682--4691, 2020.

\bibitem{Chorin_MC_1968}
A.~Chorin.
\newblock Numerical solution of the {Navier-Stokes} equations.
\newblock {\em Math. Comp.}, 22:745--762, 1968.

\bibitem{Doostmohammadi_NC_2018}
A.~Doostmohammadi, J.~Ignes-Mullol, J.~Yeomans, and F.~Sagues.
\newblock {Active nematics}.
\newblock {\em {Nature Comm.}}, {9}:{3246}, {2018}.

\bibitem{Doostmohammadi_PRL_2016}
A.~Doostmohammadi, S.~P. Thampi, and J.~M. Yeomans.
\newblock Defect-mediated morphologies in growing cell colonies.
\newblock {\em {Phys. Rev. Lett.}}, {117}:{048102}, {2016}.

\bibitem{dziuk2013finite}
G.~Dziuk and C.~M. Elliott.
\newblock {Finite element methods for surface PDEs}.
\newblock {\em Acta Numerica}, 22:289, 2013.

\bibitem{Ellis_NP_2018}
P.~W. Ellis, D.~J.~G. Pearce, Y.-W. Chang, G.~Goldsztein, L.~Giomi, and
  A.~Fernandez-Nieves.
\newblock {Curvature-induced defect unbinding and dynamics in active nematic
  toroids}.
\newblock {\em {Nature Physics}}, {14}:{85}, {2018}.

\bibitem{giomi2015geometry}
L.~Giomi.
\newblock Geometry and topology of turbulence in active nematics.
\newblock {\em Phys. Rev. X}, 5:031003, 2015.

\bibitem{Giomi_PRL_2013}
L.~Giomi, M.~J. Bowick, X.~Ma, and M.~C. Marchetti.
\newblock Defect annihilation and proliferation in active nematics.
\newblock {\em {Phys. Rev. Lett.}}, {110}:{228101}, {2013}.

\bibitem{giomi2014defect}
L.~Giomi, M.~J. Bowick, P.~Mishra, R.~Sknepnek, and M.~Cristina~Marchetti.
\newblock Defect dynamics in active nematics.
\newblock {\em Phil. Trans. Roy. Soc. A}, 372:20130365, 2014.

\bibitem{Golovatyetal_JNS_2015}
D.~Golovaty, J.~A. Montero, and P.~Sternberg.
\newblock {Dimension reduction for the Landau-de Gennes model in planar nematic
  thin films}.
\newblock {\em J. Nonlin. Sci.}, 25:1431--1451, 2015.

\bibitem{Golovatyetal_JNS_2017}
D.~Golovaty, J.~A. Montero, and P.~Sternberg.
\newblock {Dimension reduction for the Landau-de Gennes model on curved nematic
  thin films}.
\newblock {\em J. Nonlin. Sci.}, 27:1905--1932, 2017.

\bibitem{Jeseneketal_SM_2015}
D.~Jesenek, S.~Kralj, R.~Rosso, and E.~G. Virga.
\newblock {Defect unbinding on a toroidal nematic shell}.
\newblock {\em {Soft Matter}}, {11}:{2434--2444}, {2015}.

\bibitem{julicher2018hydrodynamic}
F.~J{\"u}licher, S.~W. Grill, and G.~Salbreux.
\newblock Hydrodynamic theory of active matter.
\newblock {\em Rep. Prog. Phys.}, 81:076601, 2018.

\bibitem{keber2014topology}
F.~C. Keber, E.~Loiseau, T.~Sanchez, S.~J. DeCamp, L.~Giomi, M.~J. Bowick,
  M.~C. Marchetti, Z.~Dogic, and A.~R. Bausch.
\newblock Topology and dynamics of active nematic vesicles.
\newblock {\em Science}, 345:1135--1139, 2014.

\bibitem{Kralj_SM_2011}
S.~Kralj, R.~Rosso, and E.~Virga.
\newblock {Curvature control of valence on nematic shells}.
\newblock {\em {Soft Matter}}, {7}:{670--683}, {2011}.

\bibitem{Lubenskyetal_JPII_1992}
T.~Lubensky and J.~Prost.
\newblock {Orientational order and vesicle shape}.
\newblock {\em {J. Phys. II}}, {2}:{371--382}, {1992}.

\bibitem{Mickelin_PRL_2018}
O.~Mickelin, J.~S\l{}omka, K.~J. Burns, D.~Lecoanet, G.~M. Vasil, L.~M. Faria,
  and J.~Dunkel.
\newblock Anomalous chained turbulence in actively driven flows on spheres.
\newblock {\em Phys. Rev. Lett.}, 120:164503, 2018.

\bibitem{Miura_QAM_2018}
T.-H. Miura.
\newblock {On singular limit equations for incompressible fluids in moving thin
  domains}.
\newblock {\em Quart. Appl. Math.}, 76:215--251, 2018.

\bibitem{Napoli_PRE_2020}
G.~Napoli and S.~Turzi.
\newblock {Spontaneous helical flows in active nematics lying on a cylindrical
  surface}.
\newblock {\em {Phys. Rev. E}}, {101}:{022701}, {2020}.

\bibitem{Napolietal_PRE_2012}
G.~Napoli and L.~Vergori.
\newblock Surface free energies for nematic shells.
\newblock {\em Phys. Rev. E}, 85:061701, 2012.

\bibitem{Nestler_SM_2020}
M.~Nestler, I.~Nitschke, H.~L\"owen, and A.~Voigt.
\newblock {Properties of surface Landau-de Gennes Q-tensor models}.
\newblock {\em {Soft Matter}}, {16}:{4032--4042}, {2020}.

\bibitem{Nestleretal_JNS_2017}
M.~Nestler, I.~Nitschke, S.~Praetorius, and A.~Voigt.
\newblock Orientational order on surfaces: The coupling of topology, geometry,
  and dynamics.
\newblock {\em J. Nonlin. Sci.}, 28:147--191, 2018.

\bibitem{nestler2019finite}
M.~Nestler, I.~Nitschke, and A.~Voigt.
\newblock {A finite element approach for vector-and tensor-valued surface
  PDEs}.
\newblock {\em J. Comput. Phys.}, 389:48--61, 2019.

\bibitem{Nitschke_PRSA_2018}
I.~Nitschke, M.~Nestler, S.~Praetorius, H.~L\"owen, and A.~Voigt.
\newblock {Nematic liquid crystals on curved surfaces: a thin film limit}.
\newblock {\em {Proc. Roy. Soc. A}}, {474}:{20170686}, {2018}.

\bibitem{Nitschkeetal_book_2017}
I.~Nitschke, S.~Reuther, and A.~Voigt.
\newblock Discrete exterior calculus (({EC)} for the surface {Navier-Stokes}
  equation.
\newblock In D.~Bothe and A.~Reusken, editors, {\em Transport Processes at
  Fluidic Interfaces}, pages 177--197. Springer, 2017.

\bibitem{nitschke2019hydrodynamic}
I.~Nitschke, S.~Reuther, and A.~Voigt.
\newblock Hydrodynamic interactions in polar liquid crystals on evolving
  surfaces.
\newblock {\em Phys. Rev. Fluids}, 4:044002, 2019.

\bibitem{Nitschke_PRSA_2020}
I.~Nitschke, S.~Reuther, and A.~Voigt.
\newblock Liquid crystals on deformable surfaces.
\newblock {\em Proc. Roy. Soc. A}, 476:20200313, 2020.

\bibitem{Nitschke_PAMM_2021}
I.~Nitschke, S.~Reuther, and A.~Voigt.
\newblock {Vorticity-stream function approaches are inappropriate to solve the
  surface Navier-Stokes equation on a torus}.
\newblock {\em Proc. Appl. Math. Mech.}, 1:e202000006, 2020.

\bibitem{nitschke2020observer}
I.~Nitschke and A.~Voigt.
\newblock Observer-invariant time derivatives on moving surfaces.
\newblock {\em arXiv preprint arXiv:2007.01177}, 2020.

\bibitem{Nitschkeetal_JFM_2012}
I.~Nitschke, A.~Voigt, and J.~Wensch.
\newblock A finite element approach to incompressible two-phase flow on
  manifolds.
\newblock {\em J. Fluid Mech.}, 708:418--438, 2012.

\bibitem{Novack_SIAMJMA_2018}
M.~R. Novack.
\newblock {Dimension reduction for the Landau-de Gennes model: The vanishing
  nematic correlation length limit}.
\newblock {\em SIAM J. Math. Anal.}, 50:6007--6048, 2018.

\bibitem{Pearce_NJP_2020}
D.~J.~G. Pearce.
\newblock {Defect order in active nematics on a curved surface}.
\newblock {\em {New J. Phys.}}, {22}:{063051}, {2020}.

\bibitem{Pearce_PRL_2019}
D.~J.~G. Pearce, P.~W. Ellis, A.~Fernandez-Nieves, and L.~Giomi.
\newblock {Geometrical control of active turbulence in curved topographies}.
\newblock {\em {Phys. Rev. Lett.}}, {122}:{168002}, {2019}.

\bibitem{Rank_PF_2021}
M.~Rank and A.~Voigt.
\newblock {Active flows on curved surfaces}.
\newblock {\em {Phys. Fluids}}, {}:{}, {2021}.

\bibitem{Reuther_JFM_2020}
S.~Reuther, I.~Nitschke, and A.~Voigt.
\newblock {A numerical approach for fluid deformable surfaces}.
\newblock {\em {J. Fluid Mech.}}, {900}:{R8}, {2020}.

\bibitem{Reuther_PF_2018}
S.~Reuther and A.~Voigt.
\newblock {Solving the incompressible surface Navier-Stokes equation by surface
  finite elements}.
\newblock {\em {Phys. Fluids}}, {30}:{012107}, {2018}.

\bibitem{sbalzarini2005feature}
I.~F. Sbalzarini and P.~Koumoutsakos.
\newblock Feature point tracking and trajectory analysis for video imaging in
  cell biology.
\newblock {\em Journal of structural biology}, 151(2):182--195, 2005.

\bibitem{thampi2014instabilities}
S.~P. Thampi, R.~Golestanian, and J.~M. Yeomans.
\newblock Instabilities and topological defects in active nematics.
\newblock {\em EPL (Europhysics Letters)}, 105:18001, 2014.

\bibitem{Torres-Sanchez_JFM_2019}
A.~Torres-Sanchez, D.~Millan, and M.~Arroyo.
\newblock {Modelling fluid deformable surfaces with an emphasis on biological
  interfaces}.
\newblock {\em J. Fluid Mech.}, 872:218--271, 2019.

\bibitem{Veyetal_CVS_2007}
S.~Vey and A.~Voigt.
\newblock {AMDiS: Adaptive multidimensional simulations}.
\newblock {\em Comput. Vis. Sci.}, 10:57--67, 2007.

\bibitem{virga1995variational}
E.~G. Virga.
\newblock {\em Variational theories for liquid crystals}.
\newblock Chapman and Hall/CRC, 1994.

\bibitem{Witkowskietal_ACM_2015}
T.~Witkowski, S.~Ling, S.~Praetorius, and A.~Voigt.
\newblock {Software concepts and numerical algorithms for a scalable adaptive
  parallel finite element method}.
\newblock {\em Adv. Comput. Math.}, 41:1145--1177, 2015.

\bibitem{wold1987principal}
S.~Wold, K.~Esbensen, and P.~Geladi.
\newblock Principal component analysis.
\newblock {\em Chemometrics and intelligent laboratory systems}, 2(1-3):37--52,
  1987.

\bibitem{zhang2020dynamics}
Y.-H. Zhang, M.~Deserno, and Z.-C. Tu.
\newblock The dynamics of active nematic defects on the surface of a sphere.
\newblock {\em Phys. Rev. E}, 102:012607, 2020.

\end{thebibliography}

\end{document}